\documentclass[10pt]{nature}
\usepackage{geometry}
\usepackage{lineno}
\usepackage[pdfstartview=FitH,
CJKbookmarks=true,
bookmarksnumbered=true,
bookmarksopen=true,
colorlinks=true,
pdfborder=001,
linkcolor=blue,
anchorcolor=blue,
filecolor=blue,
urlcolor=blue,
citecolor=blue,
]{hyperref}
\usepackage[ruled, vlined]{algorithm2e}
\usepackage{diagbox}
\usepackage{epsfig}
\usepackage{subfigure}
\usepackage{amsmath}
\usepackage{longtable}
\usepackage{multicol}
\usepackage{color}
\usepackage{subfigure}
\usepackage{epstopdf}
\usepackage{bm}
\usepackage[font={scriptsize}]{caption}
\DeclareCaptionLabelFormat{mylabel}{#1 #2}
\renewcommand{\vec}[1]{\mathbf{#1}}

\begin{document}

\title{Single-cell entropy to quantify the cellular transcriptome from single-cell RNA-seq data}

\author{Jingxin Liu$^1$, You Song$^1$, Jinzhi Lei$^2$}

\maketitle
\thispagestyle{empty}

\begin{affiliations}
 \item School of Software, Beihang University, Beijing 100191, China.
 \item Zhou Pei-Yuan Center for Applied Mathematics, MOE Key Laboratory of Bioinformatics, Tsinghua University, Beijing 100084, China.
\end{affiliations}

\footnotesize
\setlength{\parskip}{0pt}
\setlength{\parindent}{0.2in}
\begin{abstract}
We present the use of single-cell entropy (scEntropy) to measure the order of the cellular transcriptome profile from single-cell RNA-seq data, which leads to a method of unsupervised cell type classification through scEntropy followed by the Gaussian mixture model (scEGMM). scEntropy is straightforward in defining an intrinsic transcriptional state of a cell. scEGMM is a coherent method of cell type classification that includes no parameters and no clustering; however, it is comparable to existing machine learning-based methods in benchmarking studies and facilitates biological interpretation.
\end{abstract}

\begin{figure*}[t]
\centering
\includegraphics[width=14cm]{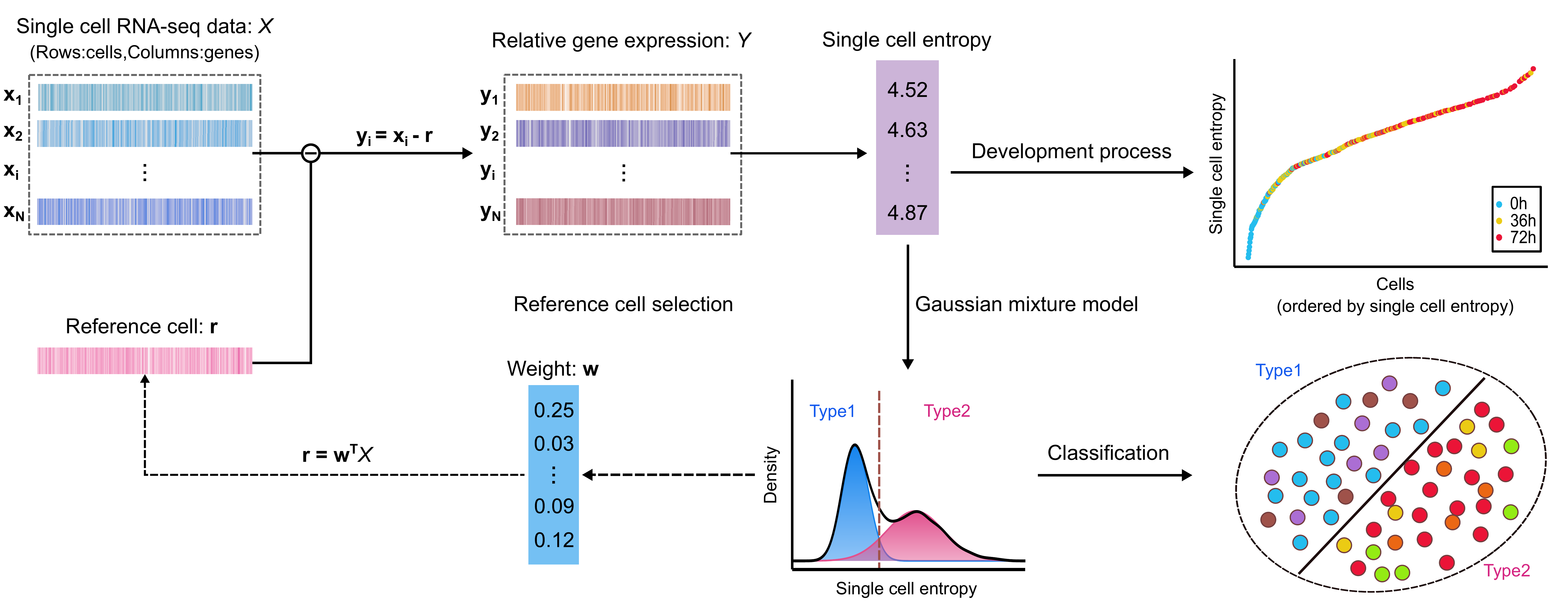}
\caption{\textbf{Overview of single-cell entropy (scEntropy) calculation, scEGMM cell type classification framework, and reference cell selection algorithm (RCSA).} Given a gene expression matrix $X$ and a gene expression vector $\vec{r}$ for the predefined reference cell, scEntropy of each cell is defined as the information entropy of the relative gene expression with respect to the reference cell. scEntropy measures the order of cellular transcription, which can be used to represent the cell differentiation process at the single-cell level (upper right panel), or to perform cell type classification based on distribution decomposition through the Gaussian mixture model (GMM) (lower right panel). We can preliminarily define the reference cell according to specific biological problems. Alternatively, we can also calculate the reference expression vector as a weight average of all cell transcriptions, with the weight given by the lowest Gaussian component of the distribution of scEntropies (dashed line arrows, see Online Methods for details). RCSA is an iterative scheme used to obtain the reference cell expression vector. The pipeline of scEntropy with reference cell selection and GMM classification provides a coherent method (scEGMM) of cell type classification. }
\label{fig:1}
\end{figure*}

\begin{multicols}{2}
Single-cell RNA sequencing (scRNA-seq) methods have allowed for the investigation of the cellular transcriptome at the level of individual cells, which gives the microscopic state of gene transcriptions. How can we define the macroscopic state of a cell based on this microscopic information? This fundamental question is essential for better interpreting the scRNA-seq data and further identifying cell types and unsupervised clustering\cite{Ren:2018jv}. Many methods of feature selection or dimensionality reduction have been developed for the interpretation and visualization of single-cell data. Commonly used methods include principal component analysis (PCA), multidimensional scaling (MDS), t-distributed stochastic neighbour embedding (t-SNE), and many other clustering approaches for the identification of cell types.%\cite{patel2014single, tirosh2016dissecting, hu2017dissecting, li2017classifying, rheaume2018single, wu2017detecting,Wang:2017jl,Wang:2018em,teschendorff2017single,Kiselev:2017jc,guo2017slice}. 
There are limitations involved in the existing methods, including the requirement of user-defined parameters, high complexity, and uncertainty of the results. There is no strong consensus about the best approach of unsupervised clustering or definition of cell types based on scRNA-seq data\cite{Kiselev:2019bk}. Here, we present single-cell entropy (scEntropy) as a cellular order parameter defined from scRNA-seq data (see \textbf{Fig. \ref{fig:1}},  Online Methods). scEntropy measures the order of cellular transcriptome with respect to a reference level; larger entropy means lower order in the transcriptions.

\begin{figure*}[t]
\centering
\includegraphics[width=12cm]{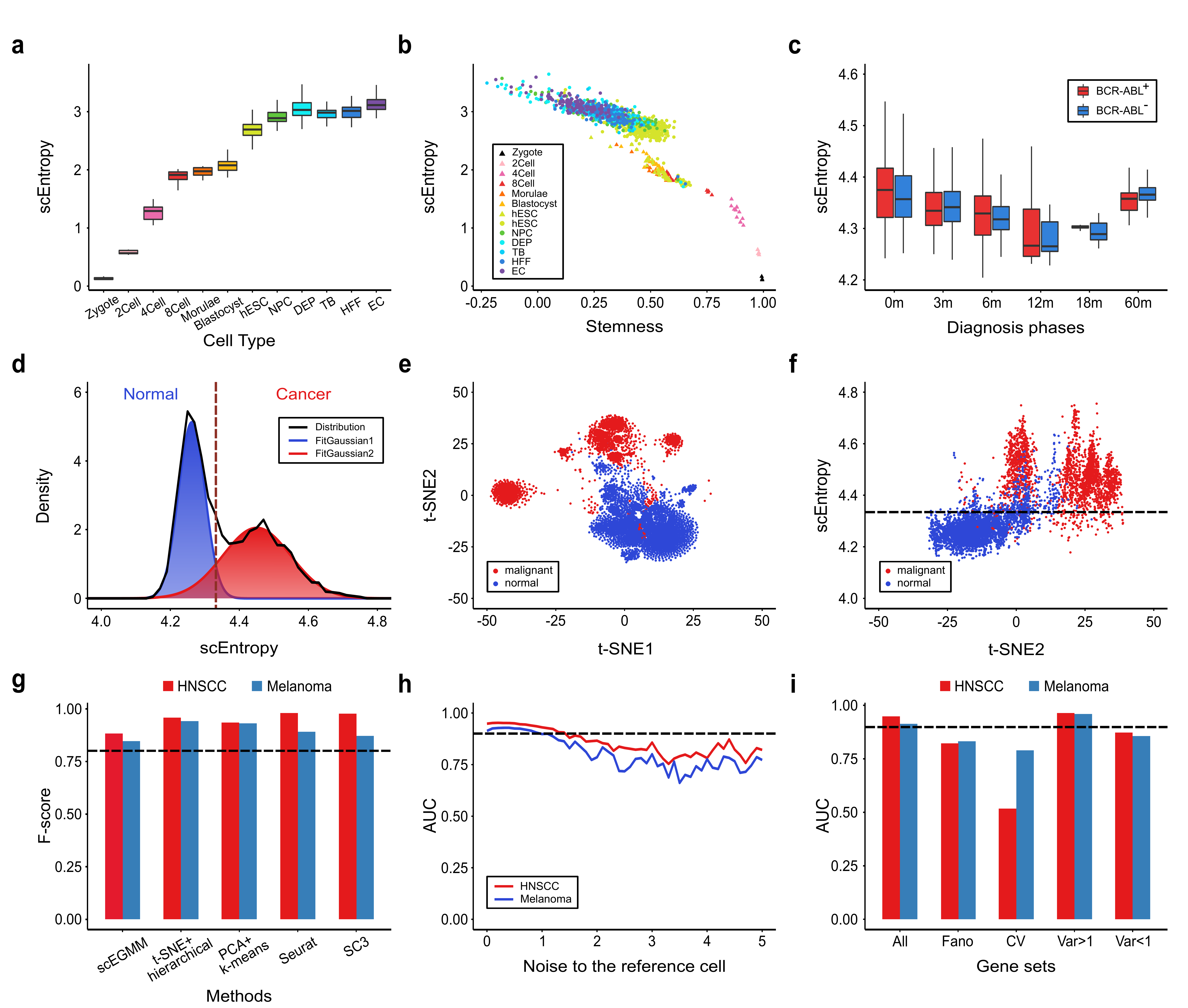}
\caption{\textbf{Applications of scEntropy and benchmark results of the scEGMM.} \textbf{a.} Using scEntropy to measure the development process of  human preimplantation embryo and human embryonic stem cell differentiation (2Cell, 2-cell embryo; 4Cell, 4-cell embryo; 8Cell, 8-cell embryo; Late, Late blastocyst; hESC, human embryonic stem cells; NPC, neural progenitor cells; DEP, definite endoderm progenitors; TB, trophoblast cells; HFF, human foreskin fibroblasts; EC, endothelial cells). \textbf{b.} Correlation between scEntropy and OCLR stemness for different cell types from 2-cell embryos to endothelial cells in early human embryos. \textbf{c.} Evolution of scEntropy of cancer stem cells (BCR-ABL$^{+}$ and BCR-ABL$^{-}$ cells) from CML patients during TKI treatment from 0 month to 60 months. \textbf{d.} GMM decomposition of the scEntropy distribution of all cells in the HNSCC dataset. The reference cell was defined as the average expression level of all normal cells (identified by CNV-based method). The vertical dashed line is the boundary to distinguish normal cells (blue) from malignant cells (red). \textbf{e.} t-SNE visualization plot for the HNSCC dataset. Different colours indicate different cell types obtained by the CNV-based method. \textbf{f.} Scatter plot of scEntropy versus t-SNE2 for HNSCC cells. The horizontal line shows the classification decision boundary of scEntropy (refer \textbf{d}). Colours indicate the cell type obtained by the CNV-based method. \textbf{g.} Benchmark results of scEGMM and  existing machine learning-based methods (t-SNE + hierarchical clustering, PCA + k-means clustering, FindCluster function in Seurat software, and SC3 clustering) on two datasets. The black dashed line indicates F-score = 0.8. \textbf{h.} Robustness of scEGMM with respect to Gaussian noise perturbations to the reference cell vector, represented by the dependence of AUC on noise strength. Black dashed line indicates AUC = 0.9. \textbf{i.} Benchmark results of scEGMM with different gene sets (All genes, highly variable genes identified by Fano index or coefficient of variation (CV), genes with expression variance $\mathrm{Var} > 1$ or expression variance $\mathrm{Var} < 1$ among all cells). The black dashed line indicates AUC = 0.9. See Online Methods for details.}
\label{fig:2}
\end{figure*}

The proposed scEntropy of a cell is defined as the information entropy of the difference in transcriptions between the cell and a predefined \textit{reference cell} (RC). The definition is straightforward and parameter free, with the RC as the only degree of freedom. Since scEntropy measures the order of the cellular transcriptome, the RC is the baseline transcriptome with the minimum entropy of zero and hence can be defined preliminarily according to specific biological problems.

scEntropy defines a macroscopic state of a cell that can be used to quantify the development process. We considered a dataset of human preimplantation embryo cells\cite{yan2013single} and a dataset of human embryonic stem cell-derived lineage-specific progenitors\cite{chu2016single}. It is straightforward to select the zygote as the RC (Online Methods). The resulting scEntropies clearly increase in preimplantation development and the subsequent differentiation process from human embryonic stem cells (hESCs) to endothelial cells (\textbf{Fig. \ref{fig:2}a}), which indicates a decrease in transcriptional order during human preimplantation and early embryo development. Moreover, we compared the scEntropy with the stemness index obtained based on the one class logistic regression (OCLR) machine learning algorithm\cite{sokolov2016one}. The OCLR stemness decreased with the scEntropy (\textbf{Fig. \ref{fig:2}b}), which indicates that the proposed scEntropy can be a measurement of cell stemness in human preimplantation and hESC differentiation.

scEntropy can also be used to track the progress of cancer development. We considered a dataset of single-cell transcriptomics of cancer stem cells from patients with chronic myeloid leukaemia (CML) throughout the disease course treated by tyrosine kinase inhibitor (TKI)\cite{giustacchini2017single}. We took the average of normal haematopoietic stem cells (HSCs) as the RC (Online Methods). The obtained scEntropy of CML stem cells (CML-SCs) decreased 12 months after TKI therapy, but increased with prolonged TKI therapy, which reveals the likely recurrence of therapy-resistant SCs (\textbf{Fig. \ref{fig:2}c}).

scEntropy provides a quantification of transcriptomic order for single cells that can distinguish malignant cells from normal tissue cells. We considered a dataset of single-cell transcriptomes from 18 patients with head and neck squamous cell carcinoma (HNSCC)\cite{puram2017single}. The dataset contains 5902 cells from different patients, which were classified into malignant and normal cells based on a copy-number variation (CNV) based method. To calculate the scEntropy of malignant cells with respect to normal cells, we took the average of all normal cell transcriptions as the RC (Online Methods). The resulting scEntropies for all cells showed a bimodal distribution with subpopulations of high or low entropies, which indicates the existence of two types of cells (\textbf{Fig. \ref{fig:2}d}). To identify the two types of cells, we applied the Gaussian mixture model (GMM)\cite{yeung2001model} to decompose the distribution of cell scEntropies, and obtained a classification that distinguishes malignant cells (high scEntropy) from non-malignant cells (low scEntropy) (see Online Methods, \textbf{Fig. \ref{fig:2}d}). The classification is consistent with that obtained from the CNV-based method (\textbf{Supplementary Table S1}). We further compared scEntropy with t-SNE analysis for each cell (Online Methods, \textbf{Fig. \ref{fig:2}e}); the scEntropy showed a positive correlation with the component t-SNE2 (\textbf{Fig. \ref{fig:2}f}). Hence, scEntropy can be considered as a one-dimensional stochastic neighbour embedding of the original data. These results show that scEntropy is a valuable method for cell type classification and dimension reduction based on scRNA-seq data.

To further demonstrate the effects of RC in cell type classification, we defined RC alternatively as the average transcriptomes of either malignant cells or all cells in the dataset. The resulting scEntropies were nearly Gaussian distribution and unable to distinguish malignant cells from normal cells (\textbf{Supplementary Fig. S1}). Hence, the selection of proper RC is essential for cell type classification with scEntropy.

To define a practicable RC, we developed a reference cell selection algorithm (RCSA) to calculate the RC iteratively from a dataset of single-cell transcriptomes (see Online Methods, \textbf{Fig. \ref{fig:1}}). The RCSA includes no parameters and gives an \textit{intrinsic reference cell} (IRC) to all cells in the dataset. The resulting IRC is a weight average of all cell transcriptomes, and the iterative scheme ensures the stability of weights so that the IRC is well defined from single-cell transcriptomes. The pipeline of scEntropy with IRC followed by GMM (scEGMM) provides a coherent method of cell type classification, which is simple and includes no parameters and no clustering (Online Methods, \textbf{Fig. \ref{fig:1}}). Moreover, the results are easy for biological interpretation, since scEntropy represents the order of the cellular transcriptome with respect to the IRC.

To benchmark scEGMM, we tested two datasets for cancer cells of HNSCC\cite{puram2017single} and melanoma\cite{tirosh2016dissecting}, and considered four other machine learning-based methods: t-SNE followed by hierarchical clustering\cite{venteicher2017decoupling}, PCA followed by k-means clustering\cite{Ding:2004dv}, shared nearest neighbour (SNN) graph-based clustering algorithm\cite{waltman2013smart}, and single-cell consensus clustering (SC3)\cite{Kiselev:2017jc}. We compared the ability of identifying malignant cells from normal cells through an F-score (see Online Methods). scEGMM is comparable to the machine learning-based methods in the two datasets (\textbf{Fig. \ref{fig:2}g}), with the F-score values larger than $0.8$ but largely reduces the uncertainty and complexity.

To show the robustness of cell type classification with the reference cell, we performed random perturbations to the IRC and evaluated the performance of cell type classification in the above two datasets by comparing the area under curve (AUC) value of the receiver operating characteristic (ROC) (see Online Methods, \textbf{Supplementary Fig. S2}). The AUC values for both datasets are insensitive to the perturbation strength and remain larger than $0.9$ when the perturbation variance is less than $1$ (\textbf{Fig. \ref{fig:2}h}). Hence, the cell type classification based on scEntropy is robust with respect to the reference cell.

scEntropy is the information entropy of transcriptions with respect to RC for all genes. We asked whether it is possible to select a subset of genes that can speed up the calculation and filter out the unnecessary genes. To this end, we applied scEGMM to the above two cancer cell datasets with five gene sets: all genes, highly variable gene sets selected by either Fano index\cite{giustacchini2017single} or coefficient of variation (CV)\cite{brennecke2013accounting}, and genes selected by either high ($\mathrm{Var > 1}$) or low ($\mathrm{Var < 1}$) variance among all cells (Online Methods). Different gene sets may affect the scEntropy values (\textbf{Supplementary Fig. S3}). The AUC values of scEGMM cell type classification based on the gene sets are shown in \textbf{Fig. \ref{fig:2}i}. The result based on the subset genes with larger variance ($\mathrm{Var > 1}$) outperforms the result based on all genes, while the other three subsets (Fano, CV, and $\mathrm{Var<1}$) yield smaller AUC than the all genes based classification. Hence, we can filter out the low variance genes by applying scEGMM for cell type classification.

In this brief communication, we present a method to define the entropy of cell transcriptome based on scRNA-seq data. The proposed scEntropy measures the intrinsic order of the cellular transcriptome with respect to a predefined reference cell. The reference cell can either be specified in accordance with the biological problem or generated from a dataset through the proposed reference cell selection algorithm. The concept of entropy has been introduced in many methods of scRNA-seq data analysis; however, all existing methods rely on other information, such as clustering of cells or genes \cite{grun2016novo,guo2017slice}, or protein-protein interaction (PPI) networks\cite{teschendorff2017single}, and hence may introduce extra uncertainty and complexity. scEntropy followed by the Gaussian mixture model provides a method of cell type classification (scEGMM) based on the bimodal/multimodal distribution of the entropies of cells. This classification method is computationally simple, includes no parameters and no clustering, is comparable to benchmark machine learning-based methods, and facilitates biological interpretation; hence, it may provide a possible solution for the challenges in unsupervised clustering of single-cell RNA-seq data\cite{Kiselev:2019bk}.

The proposed scEntropy ignores the expression of specific genes and overlooks the macroscopic cellular state through the overall changes in transcriptome with respect to a reference cell. This logic conceptually highlights the macroscopic feature of a cell, which is different from the traditional ideas of looking for biomarker differences in cell type classification. Finally, the idea of finding the entropy of a system with reference to a baseline can easily be generalized to other applications, which extends the classical concepts of entropy in describing complex systems.

\section*{METHODS}
Methods, including statements of data availability and any associated accession codes and references, are available in the online version of the paper.

\section*{ACKNOWLEDGMENTS}
The authors thank H. Ren and L. Hilbert for helpful discussions. This research is funded by the National Natural Science Foundation of China (NSFC91730101 and 11831015).

\section*{AUTHOR CONTRIBUTIONS}
J. Lei conceived the study; J. Liu performed the study and developed the software; J. Liu and Y. Song contributed to the data analysis method; J. Lei and Y. Song supervised the research; J.Lei and J. Liu lead the writing of the manuscript with input from Y. Song. All authors read and approved the final manuscript.

\section*{COMPETING FINANCIAL INTERESTS}
The authors declare no competing financial interests.

\section*{References}
\vspace{48pt}

{
\scriptsize
\setlength{\parskip}{0pt}
%\bibliographystyle{naturemag}
%\bibliography{paper}

}

\newpage
\section*{METHODS}
{

\subsection{Single-cell entropy (scEntropy).} Given an $N\times M$ gene expression matrix $X$ with $N$ cells and $M$ genes as an input, and the gene expression vector $\vec{r}$ of the reference cell. Let $\vec{x}_i\ (i=1,\cdots, N)$ the gene expression vector of the $i^{\mathrm{th}}$ cell. Calculation of the scEntropy of $\vec{x}_i$ with reference to $\vec{r}$, $S(\vec{x}_i | \vec{r})$, includes two steps:
\begin{enumerate}
\item Calculate the difference between $\vec{x}_i$ and $\vec{r}$
\begin{equation}
\vec{y}_i = \vec{x}_i - \vec{r} = (y_{i1}, y_{i2},\cdots, y_{iM}).
\end{equation}
\item The entropy $S(\vec{x}_i | \vec{r})$ is defined as the information entropy of the signal sequence $\vec{y}_i$, \textit{i.e.},
\begin{equation}
S(\vec{x}_i | \vec{r}) = - \int p_i(y) \ln p_i(y) dy,
\end{equation}
where $p_i(y)$ is the distribution density of the components $y_{ij}$ in $\vec{y}_{i}=\vec{x}_i - \vec{r}$.
\end{enumerate}

\subsection{The Gaussian mixture model (GMM) decomposition and cell type classification.}
Given the scEntropies $s_i = S(\vec{x}_i | \vec{r})\ (i=1,\cdots, N)$ for all cells in a dataset $X$, it is easy to obtain the probability density function $f(s)$ of all scEntropies.  We applied the Gaussian mixture model (GMM, as implemented in the \textit{mcluster} R package and \textit{sklearn} Python package) to decompose the distribution $f(s)$ into Gaussian components (\textbf{Fig. 1})
\begin{equation}
\label{eq:3}
f(s) = \sum_{j=1}^m c_j f_j(s),
\end{equation}
where $m$ is the number of subcomponents, $c_j$ are coefficients satisfy $c_j>0$ and $\sum_{j=1}^mc_j=1$, and $f_j(s)$ is the density functions of the Gaussian distribution $\mathcal{N}(\mu_j, \sigma_j^2)$. Here $\mu_1 < \mu_2 < \cdots < \mu_m$.

From \eqref{eq:3}, there are $m$ types of cells $\mathcal{C}_j\ (j=1,\cdots, m)$, each of which has mean scEntropy $\mu_j$, and the probability of a cell $\vec{x}$ in the class $\mathcal{C}_j$ is
\begin{equation}
p(\mathcal{C}_j | \vec{x}) = c_j f_j(S(\vec{x} | \vec{r})).
\end{equation}
Hence, we have a cell type classification method based on the misclassification rate minimization criterion: a cell $\vec{x}$ belongs to type $\mathcal{C}_j$ if $p(\mathcal{C}_j | \vec{x}) >  p(\mathcal{C}_i | \vec{x})$ for any $i\not=j$.
\\

\subsection{Reference cell selection algorithm (RCSA).}
Reference cell selection algorithm (RCSA) is an approach to define a reference cell $\vec{r}$ from a gene expression matrix $X$, which is given by a weight average of gene expression vectors of cells in a subpopulation with low scEntropy values. The algorithm is given below (\textbf{Fig. 1}):
\begin{enumerate}
\item Initialize a weight vector $\vec{w} = (\frac{1}{N}, \frac{1}{N},\cdots ,\frac{1}{N})$.
\item Define the initial reference cell
\begin{equation}
\vec{r} = \vec{w}^T X = \sum_{i=1}^N w_i\vec{x}_i.
\end{equation}
\item Calculate the scEntropies $S(\vec{x}_i | \vec{r})$ of all cells and the corresponding probability density $f(s)$.
\item Apply GMM to $f(s)$ to obtain the decomposition (we usually set $m=2$)
$$f(s) = \sum_{j=1}^m c_j f_j(s),\ c_j > 0, \ \sum_{j=1}^m c_j = 1,$$
where $f_j(s)$ is the density function of $\mathcal{N}(\mu_j, \sigma_j^2$) and $\mu_1 < \mu_2 < \cdots < \mu_m$.
\item Calculate the new weight $\vec{w}_{\mathrm{new}}$ with components $w_{\mathrm{new},i}$ given by
\begin{equation}
w_{\mathrm{new},i} = \dfrac{f_1(S(\vec{x}_i | \vec{r}))}{\displaystyle\sum_{j=1}^{N} f_1(S(\vec{x}_j | \vec{r}))},\quad (i=1,\cdots, N).
\end{equation}
\item If the mean square error $err = \frac{1}{N}\|\vec{w}_{\mathrm{new}} - \vec{w}\|^2 < 10^{-6}$, set $\vec{r} = \vec{w}_{\mathrm{new}}^T X$ as the resulting reference cell and stop the iteration; otherwise, let $\vec{w}=\vec{w}_{\mathrm{new}}$ and go to step 2.
\end{enumerate}

The above algorithm gives a reference cell $\vec{r}$ from the input matrix $X$. We can see that there is no user-defined parameters in the algorithm. Hence, the obtained reference cell is an intrinsic reference to the gene expression matrix and named as the \textit{intrinsic reference cell}.
\\

\subsection{Cell type classification with scEntropy followed by GMM (scEGMM).}
Given a gene expression matrix $X$ with $N$ cells, scEGMM includes two steps: (1) obtain the intrinsic reference cell $\vec{r}$ using RCSA, and (2) perform GMM cell type classification based on the scEntropies $S(\vec{x}_i | \vec{r})$ of all cells.
\\

\subsection{Benchmarking against existing methods.}
To benchmark scEGMM, we applied the method to two datasets of cancer cells in HNSCC (GSE103322) and in melanoma (GSE72056), and compared the results with existing methods: t-SNE + hierarchical, PCA + k-means, shared nearest neighbour (SNN) graph-based clustering algorithm in Seurat software, and single-cell consensus clustering (SC3).

For t-SNE + hierarchical, we used \textit{TSNE} in the sklearn Python package with default parameters, followed by \textit{AgglomerativeClustering} in sklearn with parameters \textit{n\_clusters=2}, \textit{linkage=``ward"}. For PCA + k-means, we used \textit{PCA} in the sklearn Python package with parameter \textit{n\_components = 2} followed by \textit{KMeans} in sklearn with parameter \textit{n\_clusters=2}. For Seurat, we used \textit{FindClusters} function in Seurat R package (version 2.3) with parameters \textit{reduction.type=``pca"}, \textit{dims.use=2}, and \textit{resolution=0.02} in both datasets. The parameter \textit{resolution} was set to a small value to obtain two clusters since \textit{FindClusters} function did not provide the explicit parameter for the number of clusters. For SC3, we used \textit{sc3} function in the SC3 R package with parameter \textit{ks = 2} (following the tutorial at \href{http://bioconductor.org/packages/release/bioc/vignettes/SC3/inst/doc/SC3.html}{http://bioconductor.org/packages/release/bioc/vignettes/SC3/inst\\/doc/SC3.html}). The \textit{sc3} function had a limitation on the cell numbers ($\leq 5000$). Since the HNSCC dataset included more than 5000 cells, we first used \textit{sc3} function with parameters \textit{ks=2}, \textit{biology=FALSE}, and \textit{svm\_num\_cells=5000} to obtain cell cluster types of 5000 random cells, and then applied hybrid support-vector machine (SVM) approach (\textit{sc3\_run\_svm} function in SC3 package) with parameter \textit{ks=2} to predict the cell cluster types of the remaining cells.

To evaluate the results of cell type classification, we compared the cluster types with the binary cell label vector obtained from CNV-based method\cite{puram2017single,tirosh2016dissecting}. Next, we defined an F-score to measure the classification performance
\begin{eqnarray*}
\mbox{F-score} = \frac{2 \times \mathrm{precision} \times \mathrm{recall}}{\mathrm{precison} + \mathrm{recall}} \\
\mathrm{precision} = \frac{\mathrm{TP}}{\mathrm{TP} + \mathrm{FP}} \\
\mathrm{recall} = \frac{\mathrm{TP}}{\mathrm{TP} + \mathrm{TN}}
\end{eqnarray*}
where $\mathrm{TP}, \mathrm{FP}$, and $\mathrm{TN}$ refer to true positive, false positive, and true negative cell numbers in the dataset, respectively, comparing with the CNV-based method. Larger F-score means higher performance.
\\

\subsection{Robustness with random perturbation to the reference cell.}
To test the robustness with random perturbation to the reference cell, we perturbed the intrinsic reference cell $\vec{r}$ with a vector of Gaussian random numbers $\mathcal{N}(0,\sigma)$, \textit{i.e.},
$$r_i \rightarrow r_i + \mathcal{N}(0,\sigma),$$
and performed the scEGMM classification to the two datasets of HNSCC and melanoma cancer cells. To compare the results with varying noise strength $\sigma$, we calculated the area under curve (AUC) value of the receiver operating characteristic(ROC) using \textit{auc} function in the sklearn Python package. The standard deviation $\sigma$ of Gaussian noise ranged from 0 to 5 with a step of 0.1, and AUCs of each dataset were generated to form a line chart as shown in \textbf{Fig. 2h}.
\\

\subsection{Effects of different gene sets.}
For a gene expression matrix $X$ of $N$ cells and $M$ genes, we can perform scEGMM classification with a subset of the $M$ genes. To evaluate the effect of cell type classification with different selection of gene subsets, we compared five gene sets:  all genes, highly variable genes (HVGs) identified by Fano index or coefficient of variation (CV), genes with high expression variance ($\mathrm{Var} > 1$) among all cells, or low expression variance ($\mathrm{Var} < 1$). We selected the HVGs using the \textit{FindVariableGenes} function in the Seurat R package. Fano index-based HVGs were identified by the \textit{FindVariableGenes} function with default \textit{``dispersion.function"} parameter value, and CV-based HVGs were identified by the \textit{FindVariableGenes} function with parameter \textit{``dispersion.function"} given by the CV value calculated from a custom function. We set the lower bound threshold $0.5$ to select the genes.

For the two datasets of HNSCC and melanoma cancer cells, scEntropy values were calculated based on the five gene sets. The AUC values for scEGMM classification were calculated referring to the binary cell label vector obtained by CNV-based method using \textit{auc} function in the sklearn Python package.
\\

\subsection{Analysis of scEntropy for datasets of human preimplantation embryo cells and embryonic stem cells (hESCs).}

To analyze the scEntropy dynamics of early human embryo cells from zygote to embryonic stem cells, we selected two datasets, the human preimplantation embryo cells (from zygote to hECS) (GSE36552) and human embryonic stem cells (from hESC to endothelial cells (ECs)) (GSE75748). First, we applied the deconvolution-based normalization method\cite{lun2016pooling} to each dataset, and selected the subset of HVGs through the HVG finding method\cite{brennecke2013accounting} on each dataset. Next, the normalized data were used for batch effect elimination through the mutual nearest neighbors (MNNs) based batch effect correction algorithm\cite{haghverdi2018batch} (as implemented in the scran R package).

After the batch effect correction, the average expression level of zygote cells was chosen as the reference cell $\vec{r}$ to calculate the scEntropy of each cell.

To calculate the stemness, OCLR method was implemented by the gelnet R package (following the authors' tutorial at \href{http://tcgabiolinks.fmrp.usp.br/PanCanStem/mRNAsi.html}{http://tcgabiolinks.fmrp.usp.br/PanCanStem/mRNAsi.html}). We trained the OCLR model with all zygote cells in GSE36552 and calculated stemness values of all cells in the blended datasets. The OCLR stemness was based on the spearman correlation between gene expression vector and model parameter vector; hence, the stemness value ranged from -1 to 1 (as shown in \textbf{Fig. 2b}, the stemness values of some cells were negative).
\\

\subsection{Analysis of CML dataset.}
To calculate the scEntropy of cancer stem cells in CML patients (GSE76312), we defined the reference cell $\vec{r}$ as the average transcription of all normal haematopoietic stem cells (HSCs) (included in the same dataset), and then calculate scEntropy of each cell with respect to the reference cell.
\\

\subsection{Data Availability.}
The scRNA-seq data used in this manuscript are available through the access number from GEO: GSE103322, GSE72056, GSE76312, GSE75748, GSE36552  (\href{https://www.ncbi.nlm.nih.gov/geo}{https://www.ncbi.nlm.nih.gov/geo}). The cancer cell datasets GSE103322 (HNSCC) and GSE72056 (melanoma) include CNV-based cell type classification to distinguish malignant cells and normal cells, and hence were used to benchmark our cell type classification method scEGMM.

Below are detailed descriptions of all datasets.
\begin{enumerate}
\item Malignant and normal cells in HNSCC (GSE103322)\cite{puram2017single}. This dataset contains 5902 cells from 18 HNSCC patients, which includes  2215 malignant cells and 3363 normal cells identified by CNV-based method. Cell types of the remaining 324 cells are unknown, which were not included in our analysis.
\item Malignant and normal cells in melanoma (GSE72056)\cite{tirosh2016dissecting}. This dataset contains 4645 cells from 19 melanoma patients, and 1257 malignant cells and 3256 normal cells were identified by CNV-based method. The remaining 132 cells could not be identified by the CNV-based method.
\item Cancer stem cells in CML (GSE76312)\cite{giustacchini2017single}. This dataset contains 232 normal bone marrow (NBM) cells and 2055 cancer stem cells ($\mathrm{CD34^+CD38^-}$ surface phenotype) throughout the tyrosine kinase inhibitor (TKI) treatment course (from diagnosis to 5 years TKI treatment). The NBM cells were used to define the reference cell. The cancer stem cells are marked with BCR-ABL status, BCR-ABL$^{+}$ and BCR-ABL$^{-}$, respectively.
\item Human embryonic stem cells under several differentiation stages (GSE75748)\cite{chu2016single}. This dataset combines cell differentiation data with different cell types and differentiation time. The cell type data includes 1019 cells ranging from human embryonic stem cells (hESCs) to  endothelial cells(ECs). The differentiation time data contains 758 cells with 6 differentiation time points from 0 to 96 hours starting from hESC.
\item Human preimplantation embryo cells (GSE36552)\cite{yan2013single}. This dataset contains 124 cells from human preimplantation embryos and human embryonic stem cells (Oocyte, 3 cells; Zygote, 3 cells; 2-cell embryo, 6 cells; 4-cell embryo, 12 cells; 8-cell embryo, 20 cells; Morulae, 16 cells; Late blastocyst, 30 cells; hESC, 34 cells).
\end{enumerate}
}

\subsection{Code Availability.}
The codes to calculate scEntropy and perform scEGMM cell type classification are freely available from \href{https://github.com/jzlei/scEntropy}{https://github.com/jzlei/scEntropy}.
\end{multicols}
\end{document}